\documentclass[pre,twocolumn,aps,floats,superscriptaddress,longbibliography,
floatfix,showpacs,showkeys,amsmath,amssymb,nofootinbib]{revtex4-1}

\usepackage{dcolumn}
\usepackage{bm}
\usepackage{txfonts}
\usepackage{pxfonts}
\usepackage{graphicx}
\usepackage{psfrag}
\usepackage{times}
\usepackage{multirow} 

\begin{document}

\title{Global dynamics of GDP and trade}

\author{Abhin Kakkad}\email{p19abhink@iima.ac.in; abhinkakkad@gmail.com}
\affiliation{Indian Institute of Management Ahmedabad, 
Vastrapur, Ahmedabad 380015, Gujarat, India}


\author{Arnab K. Ray}\email{arnab\_kumar@daiict.ac.in; arnab.kr311@gmail.com}
\affiliation{Dhirubhai Ambani Institute of Information and 
Communication Technology, Gandhinagar 382007, Gujarat, India}

\date{\today}

\begin{abstract}
We use the logistic equation to model the dynamics of the GDP 
and the trade of the six
countries with the highest GDP in the world, namely, 
USA, China, Japan, Germany, UK and India. From the modelling of
the economic data, which are made available by the World Bank, 
we predict the maximum values of the growth of GDP and trade, 
as well as the duration over which exponential growth can be sustained. 
We set up the correlated growth of GDP and trade as the phase 
solutions of an autonomous second-order dynamical system. 
GDP and trade are related to each other by a power law, whose 
exponent seems to differentiate the six national economies 
into two types. 
Under conducive conditions for economic growth, our conclusions 
have general validity. 
\end{abstract}


\maketitle

\section{Introduction}
\label{sec1} 
The Gross Domestic Product (henceforth abbreviated as GDP) of a 
country is the value of goods and services produced by the country 
in a prescribed span of time~\citep{mas07,gmacl07}, which 
customarily is a year. GDP thus quantifies the aggregate outcome of 
the economic activities of a country that are carried out all 
round the year.
As such, the GDP of a national economy is a dynamic quantity whose 
evolution (which commonly implies growth) can be 
followed through time. 

Contribution to the GDP of a country comes from another dynamic 
quantity --- the annual trade in which the country engages 
itself~\citep{gmacl07}.
The global trade network among countries exhibits some
typical properties of a complex network, namely, a scale-free  
degree distribution and small-world clusters~\citep{sb03}. 
If countries are to be treated as vertices in this network, then 
global trade can be viewed as the exchange of wealth among the 
vertices~\citep{gl04}.  
The fitness of a vertex (a country) is measured by 
its GDP, which also stands for the potential ability of a vertex
to grow trading relations with other vertices~\citep{gl04}. 
Moreover, GDP itself follows its own power-law 
distribution~\citep{gl04,gmacl07}, which in turn determines the topology 
of the global trade network~\citep{gl04}. 
In qualitative terms, 
these networks-based perspectives of the interrelation between GDP 
and trade are in agreement with 
the Gravity Model of trade, which mathematically formulates the trade 
between two 
countries to be proportional to the GDP of both~\citep{tin62} 
(also see~\citep{jea10,dbt11} for subsequent reviews). 
Considering all of the foregoing facts together, 
it is quite obvious that GDP and trade
are intimately correlated. Both form a coupled system, in which the 
dynamics of the one reinforces the dynamics of the other.   

In the present work, we look at the coupled dynamics of GDP and trade
within the general mathematical framework of autonomous nonlinear 
dynamical systems~\citep{stro}. The autonomous nonlinear equation 
with which we model the dynamics of GDP and trade is the logistic 
equation~\citep{stro,braun}. The temporal evolution of the total GDP 
of the world economy (measured in US dollars) from 1870 to 2000 does 
hint at a trend that may be modelled by the logistic equation~\citep{mas07}. 
Empirical evidence also exists for a power-law feature in the 
interdependent growth of GDP and trade~\cite{bmskm08}. In our work 
we construct a unifying theoretical model for these apparently 
unrelated observations, with our attention on 
countries that are ranked high globally in terms of their national 
GDP. From a macroeconomic perspective, GDP
is a standard yardstick with which the state of a national economy
is gauged, and in a global comparison of national economies, 
the GDP of a country is a reliable point of reference. By this
criterion, the top six economies that we study 
pertain to USA, China, Japan, 
Germany, UK and India. At present 
these six countries account 
for nearly 60\% of the global GDP and nearly 40\% of the global trade. 
China, India and USA are the three most populous countries of the
world, accounting for almost 40\% of the world population. On the 
scale of strategic economic regions, 
the three most dominant economies in the North-Atlantic region are
USA, Germany and UK. Likewise, the three most dominant economies in 
the Indo-Pacific region are China, Japan and India, not to mention 
the economic presence of USA in the same region as well. All six 
countries are members of important economic blocs like G7 and BRICS. 
USA, Japan, Germany and UK belong to the former bloc, while China 
and India belong to the latter. 
Besides, all of these countries are the leading global
representatives of three types of economic systems, 
namely, free economies (USA, Japan, Germany and UK), controlled 
economies (China) and mixed economies (India). 
That only six countries, closely connected among 
themselves through economic ties, should exert such an
overarching influence on the global economy is compatible with the 
scale-free degree distribution of both GDP~\citep{gl04,gmacl07} 
and trade~\citep{sb03}, with, additionally, a small-world cluster 
for trade networks~\citep{sb03}. 
These features would not be qualitatively altered if more countries 
were to be included in our survey. For example, G20, which is an 
economic bloc comprising the European Union and nineteen 
independent countries (including the six that we consider here),
accounts for 80\% of the global GDP, 75\% of the global 
trade and 60\% of the world population. The 
disproportionate dominance of a few elements is the 
hallmark of a large class of 
scale-free distributions~\citep{albar02},  
to which the global economic order 
can be no exception. Summing up all of these facts, 
we argue that our study of the six countries with the 
highest GDP in the world adequately captures the essence 
of the global dynamics of GDP and trade. 

Country-wise annual data, on which we have based our modelling 
and analysis, have been collected from the World Bank website for 
both GDP~\citep{usgd,cngd,jpgd,degd,ukgd,ingd} and
trade~\citep{ustd,cntd,jptd,detd,uktd,intd}. For all the six countries,  
GDP and trade are universally measured 
in terms of US dollars. With regard to USA, 
China, Japan, UK and India, the initial year for both sets of data  
is 1960. For Germany, however, the data sets begin from 
1970. All data sets end either in 2019 or 2020. Hence, our study 
ranges over six decades in all cases but one. Since our modelling 
of the economic data is based on the logistic equation, its general
mathematical theory is first laid out in Sec.~\ref{sec2}. Thereafter, 
in Sec.~\ref{sec3} we apply the logistic equation to model the 
dynamics of the annual GDP of all the six countries.
A similar analysis of the trade data has been carried out in 
Sec.~\ref{sec4}. From the modelling exercise, we predict the time 
scale over which GDP and trade grow exponentially, 
and also the respective limits to their growth.  
In Sec.~\ref{sec5} we interpret the various outcomes of the 
logistic model in the light of contemporary policies. 
In Sec.~\ref{sec6}, for all the six countries, we plot the correlated 
growth of GDP and trade on the phase plane of an autonomous second-order 
dynamical system~\citep{stro}. The phase solutions connect GDP and 
trade to each other by a power-law relation, which is matched with the 
country-wise data. The power-law exponent appears to distinguish
the economies of large countries (with large areas and populations)
from the economies of small ones (with small areas and populations). 
The full numerical analysis, by which we quantify our study, is
summarized in Tables~\ref{t1}~and~\ref{t2}. The conclusions of our 
analysis (in Sec.~\ref{sec7}), based on the economic data of the six 
countries with the highest GDP, are globally valid. This allows us 
to propose focussed measures for augmenting international trade and 
economic growth.  

\section{The logistic equation}
\label{sec2} 
Autonomous dynamical systems of the first order have the general 
form of $\dot{x} \equiv {\mathrm d}x/{\mathrm d}t = f(x)$ where
$x \equiv x(t)$, with $t$ being time~\citep{stro}. 
An autonomous dynamical system may be linear or nonlinear, depending 
on $f(x)$ being, respectively, a linear or a nonlinear 
function of $x$~\citep{stro}. A basic model of a nonlinear 
function is given by $f(x)=ax-bx^2$, with $a$ and $b$ being fixed
parameters. This leads to the logistic equation, 
\begin{equation} 
\label{logistic} 
\dot{x} \equiv \frac{{\mathrm d}x}{{\mathrm d}t} = f(x) = ax-bx^2, 
\end{equation} 
introduced initially to study population 
dynamics~\citep{stro,braun} and later extended to multiple problems of 
socio-economic~\citep{braun,mon78,akr10} and scientific 
interest~\citep{stro}. 

Under the initial condition of $x(0)=x_0$, and with the definition 
of $k=a/b$, the integral solution of Eq.~(\ref{logistic}) is  
\begin{equation} 
\label{sologis} 
x(t) = \frac{kx_0 e^{at}}{k+x_0(e^{at}-1)}. 
\end{equation}
From Eq.~(\ref{sologis}) we see that $x$ converges to the limiting 
value of $k$ when $t \longrightarrow \infty$. This limit is known 
as the carrying capacity in studies of population dynamics, and it 
is also a fixed point of the dynamical system~\citep{stro}. This
becomes clear when we set the fixed point 
condition $\dot{x}=f(x)=0$~\citep{stro}. 
The two fixed points that result are $x=0$ and $x=k=a/b$. 

On early time scales, when $t \ll a^{-1}$, the growth of $x$ can 
be approximated to be exponential, i.e. $x \simeq x_0 \exp(at)$.
This gives $\ln x \sim at$, which is a linear relation on a 
linear-log plot. Furthermore, we can interpret $a \simeq \dot{x}/x$
as the relative (or fractional) growth rate in the 
early exponential regime. 
However, this exponential growth is not indefinite,
and on times scales of $t \gg a^{-1}$ (or $t \longrightarrow \infty$)
there is a convergence to $x=k$. Clearly, the transition from the 
exponential regime to the saturation 
regime occurs when $t \sim a^{-1}$. This
time scale corresponds to the time 
when the nonlinear term in Eq.~(\ref{logistic})
becomes significant compared to the linear term. The precise time for 
the nonlinear effect to start asserting itself can be determined from 
the condition $\ddot{x}= f^\prime (x) \dot{x} =0$ when $\dot{x} \neq 0$,
with the prime indicating a derivative with respect to $x$. 
This requires solving $f^\prime (x) = a-2bx =0$ to get $x=a/2b =k/2$.
Using $x=k/2$ in Eq.~(\ref{sologis}) gives the nonlinear time scale as
\begin{equation} 
\label{nonlint}
t_{\mathrm{nl}} = \frac{1}{a} \ln \left(\frac{k}{x_0} -1\right), 
\end{equation} 
which, we stress again, is the maximum duration over which a robust 
exponential growth can be sustained. 
Hereafter, we shall use Eqs.~(\ref{sologis})~and~(\ref{nonlint}) to 
model the dynamics of the GDP and the trade of the six countries 
that we study here. 

\section{The Dynamics of GDP}
\label{sec3}
We quantify GDP by the variable $G \equiv G(t)$, with $G$
measured in US dollars and $t$ in years. To model
the growth of $G(t)$ with the logistic equation, as in
Eq.~(\ref{logistic}), we write
\begin{equation}
\label{gdplog} 
\dot{G} \equiv \frac{{\mathrm d}G}{{\mathrm d}t} = {\mathcal G}(G)
= \gamma_1 G - \gamma_2 G^2. 
\end{equation}
Noting that $x$, $a$ and $b$ in Eq.~(\ref{logistic}) translate,
respectively, to $G$, $\gamma_1$ and $\gamma_2$ in Eq.~(\ref{gdplog}),
we can write the integral solution of $G(t)$
in the same form as Eq.~(\ref{sologis}). It then follows that
when $t \longrightarrow \infty$, $G(t)$ converges to a limiting
value, i.e. $G \longrightarrow k_G = \gamma_1/\gamma_2$.

\begin{figure}[]
\begin{center}
\includegraphics[scale=0.65, angle=0]{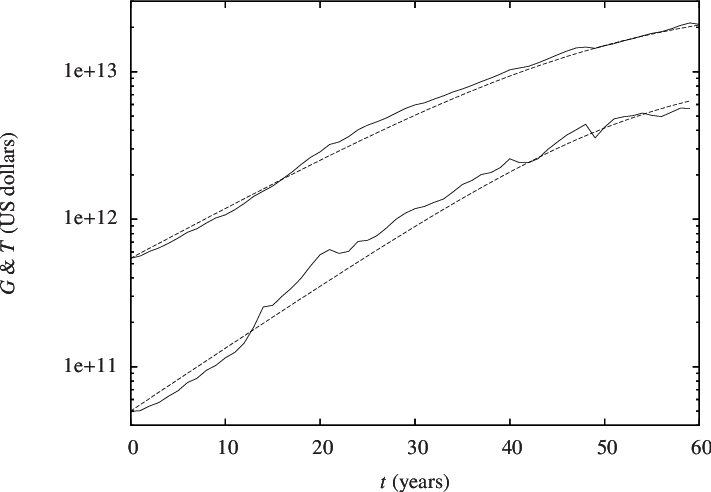}
\caption{\label{f1}\small{
Modelling the dynamics of GDP (upper plot)
and trade (lower plot) using World Bank data for USA~\citep{usgd,ustd}. 
The dotted curves follow the logistic equation with the parameter values 
in Table~\ref{t1}. The zero year of 
both plots is 1960. The GDP plot ends in 2020, but the trade plot ends
in 2019. 
}}
\end{center}
\end{figure}

\begin{figure}[]
\begin{center}
\includegraphics[scale=0.65, angle=0]{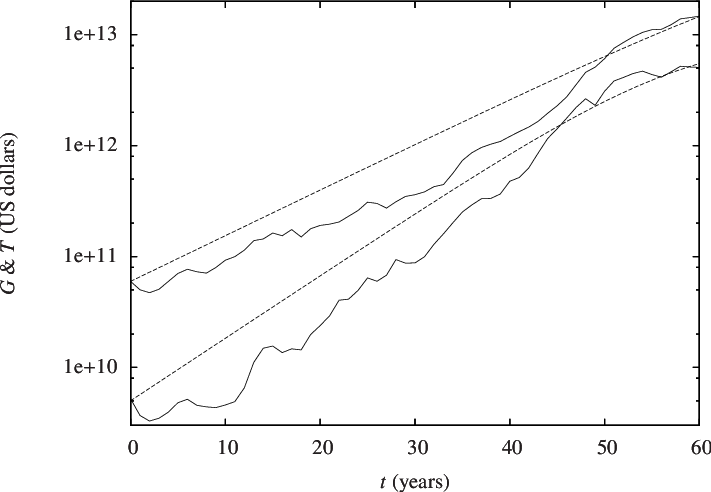}
\caption{\label{f2}\small{
Modelling the dynamics of GDP (upper plot)
and trade (lower plot) using World Bank data for 
China~\citep{cngd,cntd}.
The dotted curves follow the logistic equation with the parameter values  
in Table~\ref{t1}.
The zero year of
both plots is 1960, and both end in 2020. 
}}
\end{center}
\end{figure}

\begin{figure}[]
\begin{center}
\includegraphics[scale=0.65, angle=0]{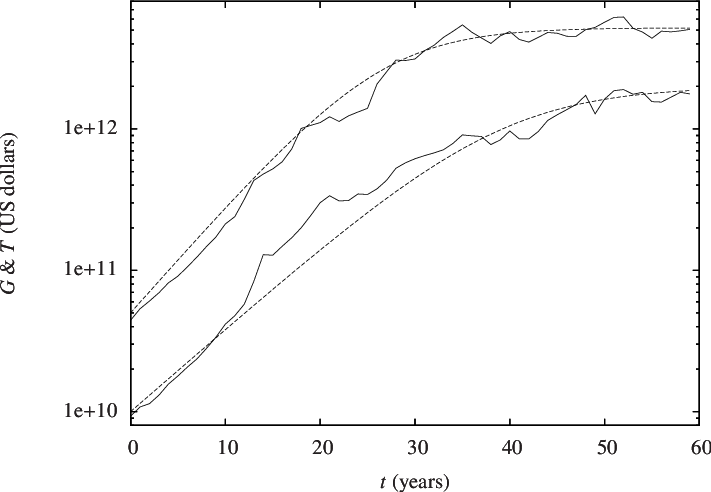}
\caption{\label{f3}\small{
Modelling the dynamics of GDP (upper plot)
and trade (lower plot) using World Bank data for
Japan~\citep{jpgd,jptd}.
The dotted curves follow the logistic equation with the parameter values  
in Table~\ref{t1}.
The zero year of both plots is 1960, and both end in 2019. 
}}
\end{center}
\end{figure}

\begin{figure}[]
\begin{center}
\includegraphics[scale=0.65, angle=0]{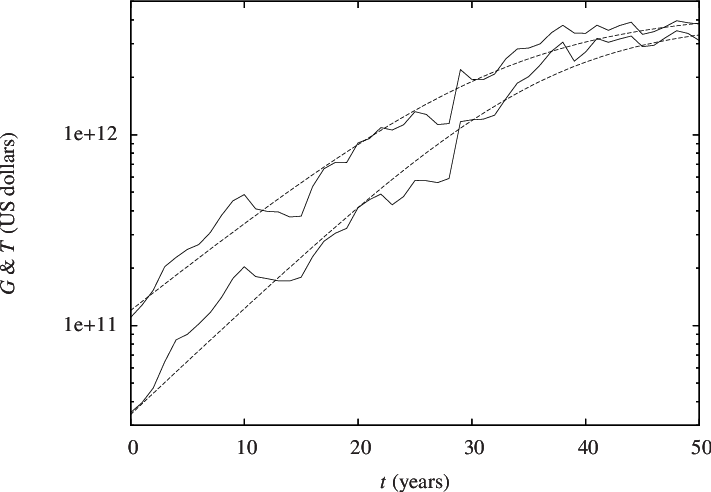}
\caption{\label{f4}\small{
Modelling the dynamics of GDP (upper plot)
and trade (lower plot) using World Bank data for
Germany~\citep{degd,detd}.
The dotted curves follow the logistic equation with the parameter values  
in Table~\ref{t1}.
The zero year of
both plots is 1970, and both end in 2020.
}}
\end{center}
\end{figure}

\begin{figure}[]
\begin{center}
\includegraphics[scale=0.65, angle=0]{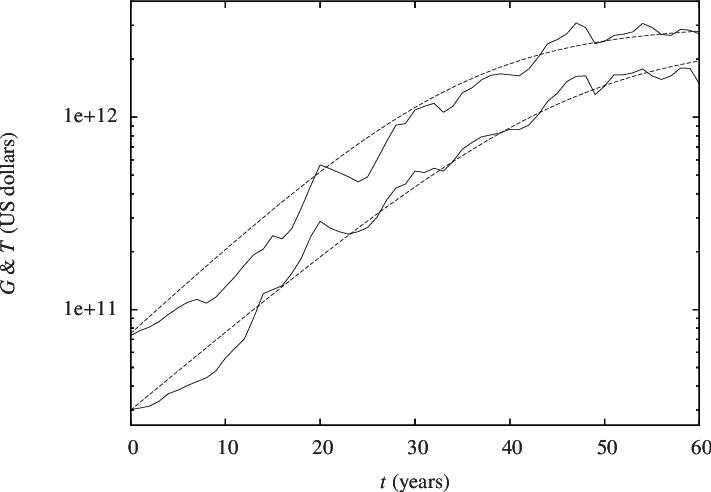}
\caption{\label{f5}\small{
Modelling the dynamics of GDP (upper plot)
and trade (lower plot) using World Bank data for
UK~\citep{ukgd,uktd}.
The dotted curves follow the logistic equation with the parameter values  
in Table~\ref{t1}.
The zero year of
both plots is 1960, and both end in 2020.
}}
\end{center}
\end{figure}

\begin{figure}[]
\begin{center}
\includegraphics[scale=0.65, angle=0]{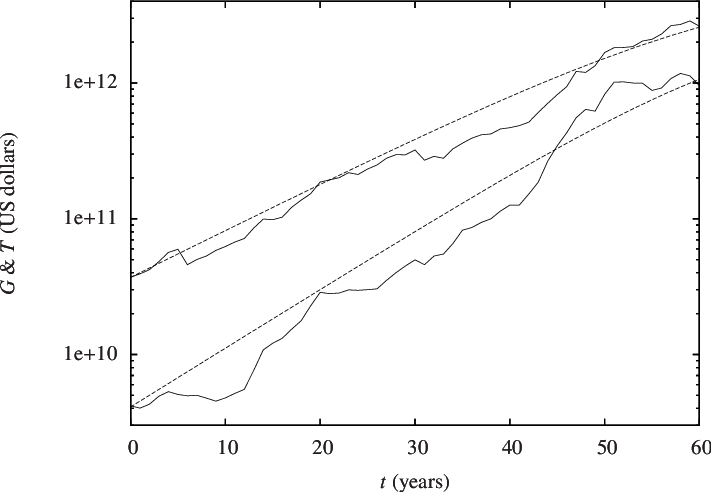}
\caption{\label{f6}\small{
Modelling the dynamics of GDP (upper plot)
and trade (lower plot) using World Bank data for
India~\citep{ingd,intd}.
The dotted curves follow the logistic equation with the parameter values  
in Table~\ref{t1}.
The zero year of
both plots is 1960, and both end in 2020.
}}
\end{center}
\end{figure}

For the six countries in our study, the early exponential
growth of the GDP and its later convergence to a finite limit are
modelled in all the
upper linear-log plots in Figs.~\ref{f1}~to~\ref{f6}.
The uneven lines follow the movement of the real GDP
data, available from the World Bank~\citep{usgd,cngd,jpgd,degd,ukgd,ingd}. 
The smooth dotted curves theoretically model the real data with the 
integral solution of Eq.~(\ref{gdplog}),
which will be in the form of Eq.~(\ref{sologis}).
The values of $\gamma_1$ (the relative annual growth rate of GDP),
$k_G$ (the predicted maximum value of GDP) and $t_{\mathrm{nl}}$
(the duration of exponential growth before the onset of nonlinearity),
calibrated through the model fitting in all the cases, are to be 
found in Table~\ref{t1}. The most
convincing match of the GDP data with the model function is seen
in Fig.~\ref{f1}, i.e. for USA. Consistent fitting of the GDP data
with the model function is also seen for Japan, Germany, UK and India,
for which Figs.~\ref{f3},~\ref{f4},~\ref{f5}~and~\ref{f6}
provide respective evidence.
Similar consistency, however, is not observed in the model fitting
of the GDP data for China, as we note from the upper plot
in Fig.~\ref{f2}. These observations about the model-fitting of the
GDP data are statistically summarized in Table~\ref{t2}, which sets
down the mean $\mu_G$ and the standard deviation $\sigma_G$ of
the yearly relative variations of the actual GDP 
data~\citep{usgd,cngd,jpgd,degd,ukgd,ingd} with respect to the 
logistic function. 

\begin{table*}[]
\caption{\label{t1} Parameter values of the logistic equation for 
dynamically
modelling the World Bank GDP data~\citep{usgd,cngd,jpgd,degd,ukgd,ingd} 
and trade data~\citep{ustd,cntd,jptd,detd,uktd,intd}
of the six countries that are listed in the first column. The 
data have been plotted and modelled in Figs.~\ref{f1}~to~\ref{f6}. 
The last column lists the values of the power-law exponent in the 
correlated growth of GDP and trade, as 
plotted and modelled in Figs.~\ref{f7}~to~\ref{f12}.
}
\begin{tabular}{|c|ccc|ccc|c|c|}

\hline
\multirow{2}{*}{} & 
\multicolumn{2}{c}{\qquad \qquad Parameters to fit $G$ (GDP)} &
\multirow{3}{*}{} & %
\multicolumn{3}{c|}{Parameters to fit $T$ (Trade)} & 
& \multirow{2}{*}{}\\

Country
& $\gamma_1$ (per annum) & $k_G$ (trillion US dollars) & 
$t_{\mathrm{nl}}$ (years) & $\tau_1$ (per annum) & 
$k_T$ (trillion US dollars) & $t_{\mathrm{nl}}$ (years) & 
$G$-$T$ Correlation & $\alpha$ \\ \hline

{USA}
& $0.080$ & $30.0$ & $50$ & $0.099$ & $10.0$ & $53$ & $0.992$ 
& $0.75$ \\

{China}
& $0.095$ & $80.0$ & $76$ & $0.130$ & $10.0$ & $58$ & $0.983$ 
& $0.65$ \\

{Japan}
& $0.175$ & $~~5.2$ & $26$ & $0.135$ & $~~2.0$ & $39$ & $0.919$ 
& $1.00$ \\

{Germany}
& $0.110$ & $~~4.4$ & $32$ & $0.130$ & $~~3.9$ & $36$ & $0.987$ 
& $0.85$ \\

{UK}
& $0.105$ & $~~3.0$ & $35$ & $0.095$ & $~~2.5$ & $46$ & $0.993$ 
& $0.90$ \\

{India}
& $0.080$ & $~~6.0$ & $64$ & $0.100$ & $~~3.0$ & $66$ & $0.982$ 
& $0.60$ \\

\hline
\end{tabular}
\end{table*}

\begin{table*}[]
\caption{\label{t2} Statistical analyses of the {\em relative} 
difference between the actual data and the model functions. 
The second column lists the mean $\mu_G$ and the standard 
deviation $\sigma_G$ of the yearly relative variations of the GDP 
data~\citep{usgd,cngd,jpgd,degd,ukgd,ingd}
with respect to the logistic model. Likewise, in the third
column, $\mu_T$ and $\sigma_T$ are the mean and the standard 
deviation, respectively, of the yearly relative variations of the 
trade data~\citep{ustd,cntd,jptd,detd,uktd,intd}. The last 
column pertains to the $G$-$T$ correlation 
(as plotted in Figs.~\ref{f7}~to~\ref{f12}), with $\mu_\alpha$ 
and $\sigma_\alpha$ being, respectively, the mean and the 
standard deviation of the yearly relative variations of the logarithm 
of the GDP data~\citep{usgd,cngd,jpgd,degd,ukgd,ingd}, with 
respect to the logarithm of the model power-law function.}
\begin{tabular}{|c|cc|cc|cc|}

\hline
\multirow{1}{*} & {Statistical analysis of $G(t)$} (GDP) &
\multirow{1}{*} & {Statistical analysis of $T(t)$} (Trade) &
\multirow{1}{*} & {Statistical analysis of $\alpha$} & \\

Country
& $\mu_G$ & $\sigma_G$ & $\mu_T$ & $\sigma_T$ & $\mu_\alpha$ & 
$\sigma_\alpha$ \\ \hline

{USA}
& $~~~0.0492$ & $0.0873$ & $~~~0.1160$ & $0.2040$ & $-0.0012$ & $0.0024$ \\

{China}
& $-0.3568$ & $0.2504$ & $-0.3570$ & $0.3393$ & $~~~0.0075$ & $0.0113$ \\

{Japan}
& $-0.0833$ & $0.1395$ & $~~~0.1900$ & $0.3682$ & $-0.0045$ & $0.0082$ \\

{Germany}
& $~~~0.0489$ & $0.1744$ & $~~~0.0736$ & $0.2411$ & $-0.0034$ & $0.0051$ \\

{UK}
& $-0.1089$ & $0.1651$ & $~~~0.0053$ & $0.1679$ & $-0.0019$ & $0.0033$ \\

{India}
& $-0.1359$ & $0.1743$ & $-0.1630$ & $0.3534$ & $~~~0.0015$ & $0.0079$ \\

\hline
\end{tabular}
\end{table*}

\section{The dynamics of trade}
\label{sec4}
The annual trade of a country accounts for the total  
import and export of goods and services.
The World Bank data on annual trade are given as a percentage of the 
annual GDP of a country~\citep{ustd,cntd,jptd,detd,uktd,intd}.
Knowing the annual GDP, the trade percentage can be 
expressed explicitly in terms of 
US dollars, which we denote by the variable $T \equiv T(t)$, 
with $t$ continuing to be measured in years.  
We model the dynamics of $T(t)$ with the logistic equation, 
as done in Eq.~(\ref{gdplog}), and write
\begin{equation}
\label{trdlog} 
\dot{T} \equiv \frac{{\mathrm d}T}{{\mathrm d}t} = {\mathcal T}(T)
= \tau_1 T - \tau_2 T^2. 
\end{equation}
Comparing Eq.~(\ref{trdlog}) with Eq.~(\ref{logistic}), we note 
that $x$, $a$ and $b$ translate, respectively, to $T$,
$\tau_1$ and $\tau_2$. Hence, from the integral solution of $T(t)$, 
which will be in the same form as Eq.~(\ref{sologis}), we will get  
a convergence of $T \longrightarrow k_T = \tau_1/\tau_2$,
when $t \longrightarrow \infty$. 

The fitting of the integral solution of Eq.~(\ref{trdlog}) with the 
trade data~\citep{ustd,cntd,jptd,detd,uktd,intd} 
is shown in all the lower linear-log plots in 
Figs.~\ref{f1}~to~\ref{f6}. 
The values of $\tau_1$ (the relative annual growth rate of trade), 
$k_T$ (the predicted upper limit of trade) and 
$t_{\mathrm{nl}}$ (when nonlinearity sets in), 
using which we fit the model equation with the data, 
are given in Table~\ref{t1}. The consistency of the model fitting 
is statistically summarized in Table~\ref{t2}, which gives the 
mean $\mu_T$ and the standard deviation $\sigma_T$ of the 
yearly relative variations of the actual  
trade data~\citep{ustd,cntd,jptd,detd,uktd,intd}
with respect to the logistic function. 
In Figs.~\ref{f1}~to~\ref{f6} we see that the model 
fitting for trade in the lower plots largely resembles the features 
of the model fitting for the GDP in the upper plots. This is very
much 
true for USA, Japan, Germany and UK on the one hand and China on
the other. In the case of India, the upper plot for GDP is more 
regular than the lower plot for trade, as can be seen in 
Fig.~\ref{f6}. 
The overall similarity between the two plots implies that 
there is a high correlation between the GDP and the trade of 
a country. Country-wise values of the GDP-trade correlation 
are in the second last column of Table~\ref{t1}. 
We look into this matter more closely in Sec.~\ref{sec6}. 

\section{Interpreting the logistic model}
\label{sec5}
The irregularity of the two plots in Fig.~\ref{f2} (and related 
values in Table~\ref{t2}) suggests that 
China is an anomalous case in modelling the dynamics of both GDP 
and trade with the logistic equation. The trade plot in Fig.~\ref{f6} 
conveys a similar hint for India. These anomalies can be explained 
from the perspective of world history in the latter half of the
twentieth century. In order to do so, we first consider all the 
three major economies of the North-Atlantic region and Japan in
the Indo-Pacific.
In the period that immediately followed the
Second World War, which ended in 1945, USA was the only country
among the principal belligerents of the war that possessed a fully
operational
industrial infrastructure. With the onset of the Cold War against
the erstwhile Soviet Union, it became a policy imperative for USA
to lend its industrial power for the economic revival of both
Western Europe (under the Marshall Plan) and Japan. This resulted
in a rapid re-industrialization of Japan and the erstwhile
West Germany.
Indeed, in the case of West Germany, the swiftness of the economic
recovery from the ravages of the war is spoken of as the ``German
economic miracle." In comparison, the post-war economic
recovery of UK, which by then had also given up many of its bountiful
colonies (most notably
India), was slow. Nevertheless, by 1960, Japan, Germany and UK
had achieved economic stability under the guidance of USA, in
consequence of which these three countries were well set on
the path of general prosperity. This comfortable state of affairs
is reflected in the
relatively ordered progression of the data (over several
decades starting from 1960) and its close match with
the model function in Figs.~\ref{f3},~\ref{f4}~and~\ref{f5}
(with support from related values in Table~\ref{t2}).

In contrast, after the Second World War,
the economic growth of China and a politically independent India
did not experience the advantages that regenerated the economies
of Japan, Germany and UK. For close to three decades after the
Second World War, China continuously suffered from internal
political upheavals like
the Great Leap Forward and the Cultural Revolution. Unsurprisingly,
therefore,
during this period the economic growth of China was severely
impeded. India, on the other hand, experienced
domestic political stability in the same period, but its benefits
were not visible on its post-colonial economic development, mainly
due to government policies. What is more, both
China and India
were in a state of war several times (once between
themselves) in the first two to three decades of their new
beginning as sovereign states.
The combined effect of all the adversities encountered by China
and India can be observed in
the irregular path traced by the GDP data in
Fig.~\ref{f2} and the trade data in Figs.~\ref{f2}~and~\ref{f6}.

In the light of the foregoing observations, 
we can now discern two distinct categories. 
In one category, the logistic equation fits the data in the expected 
manner. USA, Japan, Germany and UK 
belong to this category, with USA showing the greatest accuracy 
for the logistic fit, as in Fig.~\ref{f1} and Table~\ref{t2}. 
We note certain 
characteristics that are common to these four countries. Since the end 
of the Second World War, all of them fostered universal democratic 
values
in their internal politics, underwent no military conflict on 
their borders, and promoted free economic growth without much 
intervention from the state. The cumulative effect of these 
conditions is conducive to a natural development of material 
well-being. The absence of any one of the aforementioned 
conditions causes an imbalance and to a greater or lesser extent
creates the second category. 
China and India are in the second category. Both 
countries have a record of conflict
on their common border and borders with some of their other geographical 
neighbours. Both have government control in varying degrees on their 
respective
economies. And specific to China,  
the political conditions within the country differ from the norms
of democracy that prevail in the other five countries in this study.
Under these circumstances, economic growth 
follows an uneven course, which, in the case of China, is marked 
by the discrepancy in the
logistic modelling of the national economic data in both the
plots in Fig.~\ref{f2}, and by having the highest absolute values 
of $\mu_G$ and $\mu_T$ among all the countries in Table~\ref{t2}. 
For India, the discrepancy is partial,  
as it is mostly seen in the trade plot in Fig.~\ref{f6}. This 
mitigation is 
due to the stable democratic polity in the country. 
Going by these observations, we contend that the balanced economic 
growth of a country (especially in terms of its GDP and trade data) 
can be gauged from the closeness
of its match with the logistic equation (the closeness being
quantified by small values of $\mu_G$, $\sigma_G$, $\mu_T$ 
and $\sigma_T$ in Table~\ref{t2}). 

This raises a valid question about the general import 
of the logistic equation. It is known that the natural
growth of many systems 
is described satisfactorily by the logistic equation,  
the growth of species being a standard example~\citep{braun}.  
Hence, the logistic equation is organically compatible with 
natural evolution in an open and productive environment.   
This principle arguably applies to the free evolution of 
economic systems as well, a point of view that is  
supported by Figs.~\ref{f1},~\ref{f3},~\ref{f4}~and~\ref{f5}. 

\section{The GDP-Trade correlation}
\label{sec6} 
At the end of Sec.~\ref{sec4}, we mentioned the high correlation between 
the GDP and the trade of all the six countries in our study, something
that is evident from the values of the $G$-$T$ correlation in the 
second last column of Table~\ref{t1}. This correlation is  
expected, because GDP and trade are dynamically connected to each 
other~\citep{gl04,gmacl07,mas07}.  
As such, the coupled dynamics of GDP and trade must be governed by an 
autonomous system of the second order, given as 
${\dot T}={\mathcal T}(T,G)$ and 
${\dot G} = {\mathcal G}(T,G)$. The
$T$-$G$ phase solutions are determined by integrating 
\begin{equation} 
\label{phase} 
\frac{{\mathrm d}G}{{\mathrm d}T} = \frac{\dot{G}}{\dot{T}} = 
\frac{{\mathcal G}(T,G)}{{\mathcal T}(T,G)}
\end{equation} 
for various initial values of the $(T,G)$ coordinates~\citep{stro}. 
Since the autonomous 
functions ${\mathcal G}(T,G)$ and ${\mathcal T}(T,G)$ are not known
a priori, we proceed with a linear ansatz of 
${\mathcal G} \simeq \gamma_1 G$ from Eq.~(\ref{gdplog}) and 
${\mathcal T} \simeq \tau_1 T$ from Eq.~(\ref{trdlog}). This 
linearization is in accord with the multiplicative character of 
GDP and trade, whereby the revenue generated in one year is reinvested 
in the economic cycle of the next year~\citep{gmacl07}. 
In the linear regime, we get a scaling formula that goes as 
(with $\alpha = \gamma_1/\tau_1$)
\begin{equation} 
\label{scale} 
G(T) \sim T^{\alpha},
\end{equation} 
for which empirical evidence was found from 1948 to 2000, in a survey 
of nearly two dozen countries of varying economic strength (high, 
middle and low-income economies)~\citep{bmskm08}.\footnote{
For the coupled growth of $G$ and $T$,
a second-order dynamical system like ${\dot G} \sim T$ and 
${\dot T} \sim G$ may appear apt. This, however, gives phase solutions 
like $G^2 \sim T^2$, which is not borne out by a reported study of 
GDP and trade growth~\citep{bmskm08}.
We argue that the linear terms on the right hand sides of 
Eqs.~(\ref{gdplog})~and~(\ref{trdlog}) are the most dominant,  
and lead to Eq.~(\ref{scale}).
}

\begin{figure}[]
\begin{center}
\includegraphics[scale=0.65, angle=0]{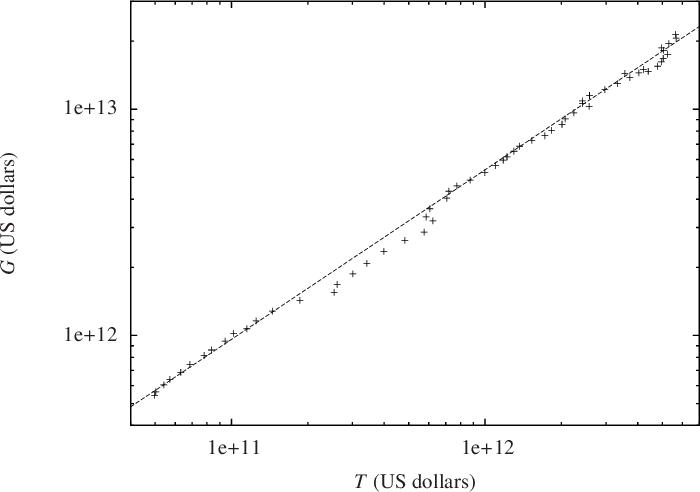}
\caption{\label{f7}\small{
Plotting GDP against trade using World Bank data for
USA~\citep{usgd,ustd}.
The dotted line follows Eq.~(\ref{scale}) with $\alpha = 0.75$
(see Table~\ref{t1}). The plot begins in 1960 and
ends in 2019.
}}
\end{center}
\end{figure}

\begin{figure}[]
\begin{center}
\includegraphics[scale=0.65, angle=0]{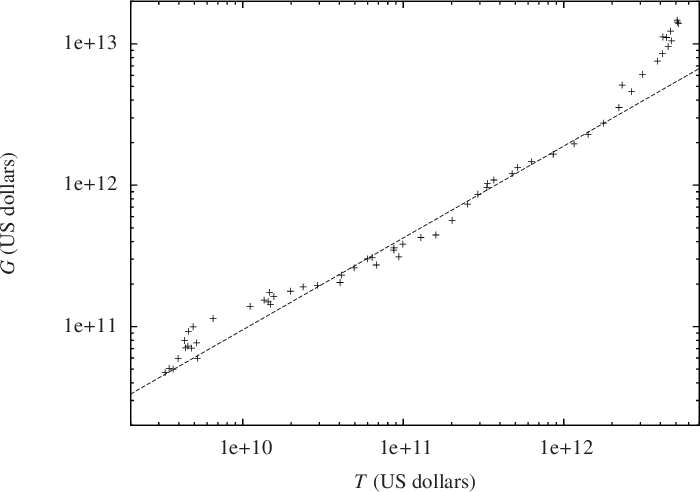}
\caption{\label{f8}\small{
Plotting GDP against trade using World Bank data for
China~\citep{cngd,cntd}.
The dotted line follows Eq.~(\ref{scale}) with $\alpha = 0.65$
(see Table~\ref{t1}). The plot begins in 1960 and
ends in 2020.
}}
\end{center}
\end{figure}

\begin{figure}[]
\begin{center}
\includegraphics[scale=0.65, angle=0]{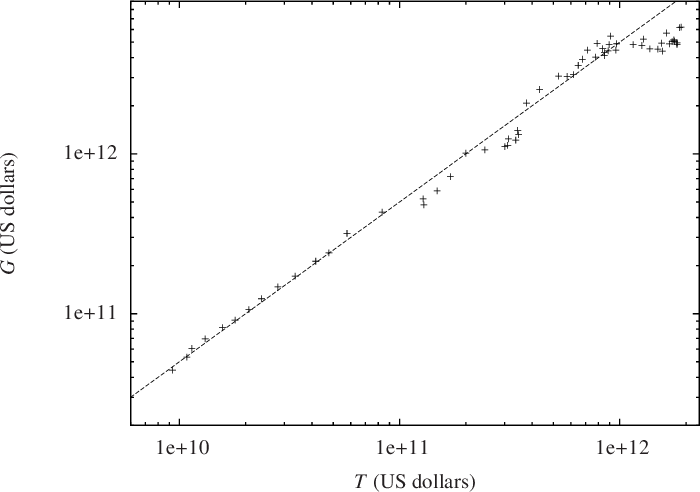}
\caption{\label{f9}\small{
Plotting GDP against trade using World Bank data for
Japan~\citep{jpgd,jptd}.
The dotted line follows Eq.~(\ref{scale}) with $\alpha = 1.00$
(see Table~\ref{t1}). The plot begins in 1960 and
ends in 2019.
}}
\end{center}
\end{figure}

\begin{figure}[]
\begin{center}
\includegraphics[scale=0.65, angle=0]{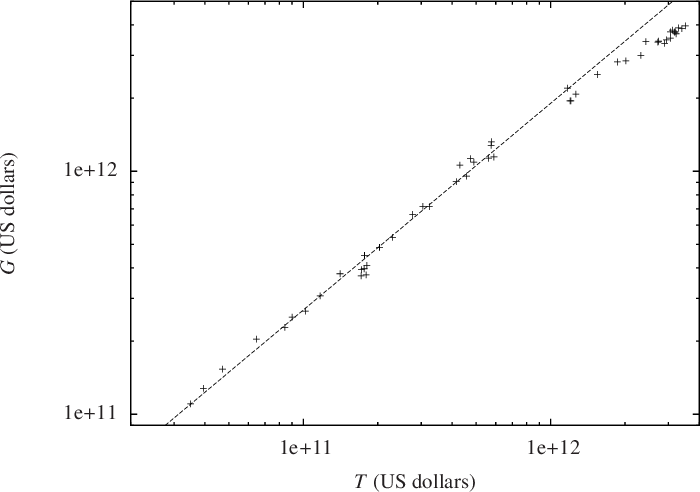}
\caption{\label{f10}\small{
Plotting GDP against trade using World Bank data for
Germany~\citep{degd,detd}.
The dotted line follows Eq.~(\ref{scale}) with $\alpha = 0.85$
(see Table~\ref{t1}). The plot begins in 1970 and
ends in 2020.
}}
\end{center}
\end{figure}

\begin{figure}[]
\begin{center}
\includegraphics[scale=0.65, angle=0]{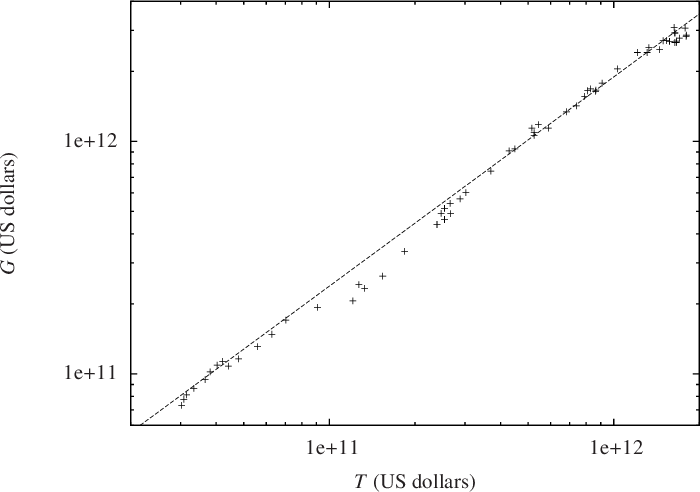}
\caption{\label{f11}\small{
Plotting GDP against trade using World Bank data for
UK~\citep{ukgd,uktd}.
The dotted line follows Eq.~(\ref{scale}) with $\alpha = 0.90$
(see Table~\ref{t1}). The plot begins in 1960 and
ends in 2020.
}}
\end{center}
\end{figure}

\begin{figure}[]
\begin{center}
\includegraphics[scale=0.65, angle=0]{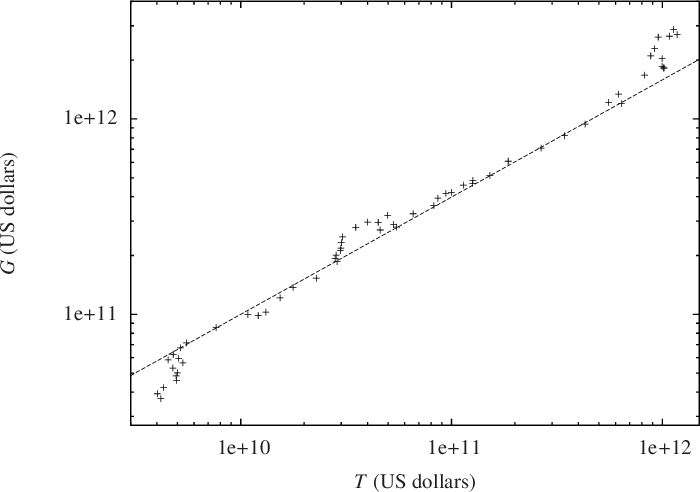}
\caption{\label{f12}\small{
Plotting GDP against trade using World Bank data for
India~\citep{ingd,intd}.
The dotted line follows Eq.~(\ref{scale}) with $\alpha = 0.60$
(see Table~\ref{t1}). The plot begins in 1960 and
ends in 2020.
}}
\end{center}
\end{figure}

The power-law function in Eq.~(\ref{scale}) becomes linear in a log-log
plot. This is indeed
what we see in Figs.~\ref{f7}~to~\ref{f12}, all of which model the 
coupled growth of GDP and 
trade up to 2020 (or 2019) for the six countries in our study.
The power-law exponent $\alpha$, which is the slope of the linear fit,
is confined within a narrow range of $0.6< \alpha <1$ in all cases. The
country-wise values of $\alpha$ have been set
down in the last column of Table~\ref{t1}. Keeping only the linear
terms in Eqs.~(\ref{gdplog})~and~(\ref{trdlog}), which gives the
phase solutions in Eq.~(\ref{scale}),
we find that $\alpha = \gamma_1/\tau_1$. The values of
$\gamma_1$, $\tau_1$ and $\alpha$ in Table~\ref{t1} do show that
$\alpha$ is practically quite close to $\gamma_1/\tau_1$. This
independently validates our modelling of GDP and trade growth with
the logistic equation.

Looking at Figs.~\ref{f7}~to~\ref{f12}, we realize that the power-law
scaling of $G$ with respect to $T$ holds true for every country over 
at least two orders of magnitude, an observation borne out consistently 
by the low values of $\mu_\alpha$ and $\sigma_\alpha$ in 
Table~\ref{t2}. For high values of $T$ and $G$,
deviation from this scaling behaviour occurs for China, Japan,
Germany and India (Figs.~\ref{f8},~\ref{f9}~\ref{f10}~and~\ref{f12},
respectively). This deviation is due to the nonlinear effects in the 
real data, which we have not considered in the coupled autonomous 
functions ${\mathcal G}(T,G)$ and ${\mathcal T}(T,G)$, 
in deriving Eq.~(\ref{scale}). We also note that
${\mathrm d}^2 G/{\mathrm d}T^2 < 0$ for $\alpha < 1$, i.e. $G$ 
increases with $T$ at a decreasing rate as time progresses. 
This explains the reduction of
the gap between the GDP and the trade plots in Figs.~\ref{f1}~to~\ref{f6}
on long time scales.

The values of $\alpha$ in Table~\ref{t1} hint at a possible distinction
between two types of countries. In one type, which includes Japan,
Germany and UK, $\alpha$ has a relatively high value. In the other
type, which includes USA, China and India, $\alpha$ has a smaller value.
The latter type of countries are geographically extended on continental
or subcontinental scales, and have large populations. 
In contrast, countries of the former type are territorially restricted
with comparatively small populations. This distinction
between the two types of countries may cause qualitative
differences in trading patterns~\citep{fr99}, with a concomitant
effect on the GDP. We conjecture that the value of $\alpha$ 
segregates the two types of countries (about 
$\alpha =0.80$). More clarity on this point, however, 
requires a wider study. 

\section{Conclusions}
\label{sec7} 
Our study has brought two salient results to the fore. The first is 
that under conducive conditions, the logistic equation suffices to 
model the growth of GDP and trade. The conducive conditions refer 
to the state of internal politics, military engagements 
and economic policies of a 
country. 
The second 
result is a correlated growth of GDP and trade, driven by a power-law. 
The power law can be traced back to the logistic equation itself and
the exponent of the power law possibly characterizes economies on 
the basis of geographical scales and population sizes.  
These theoretical 
claims are founded on empirical facts, and hold true across countries. 
Global validity can be attributed to these principles, even though our
study covers six countries, because the 
country-wise distributions of GDP and trade have a scale-free 
order~\citep{gmacl07,sb03,gl04}. 

A scale-free order, in which a small number of countries account 
for a large portion of the international trade~\citep{bmskm08}, can 
be exploited to devise globally-coordinated strategies for the 
recovery of the world economy from the current Covid-19 pandemic. The 
first step in this respect is a vigorous re-activation of the 
international trade network. For this a leading role is essential
from the two strategic economic regions that we have considered in our 
study, i.e. the North-Atlantic and the Indo-Pacific. 
Major shipping routes pass through both  
and many countries in or abutting the two oceanic regions
have high national GDP. 
These countries form a regional trading cluster, in which
their geographical proximity promotes
trade~\citep{tin62,jea10,dbt11}. Once trade 
flourishes within an economic region, its main economic players 
can then trade with other economic regions, as it ought to be 
in the scale-free and small-world architecture of the 
global trade network~\citep{sb03}. Since GDP and trade are correlated, 
enhancement of trading activities will have a positive impact on the 
GDP of the participating countries. 
     
Our theoretical modelling, based on the logistic equation, predicts
long-term economic stagnation. 
Reasons for this are dwindling natural resources, natural calamities, 
pandemics, obsolescence of technology, military conflicts, etc. The
decisive reasons are often unforeseen. Nevertheless, the logistic 
equation continues to be a favoured mathematical tool for modelling 
the evolution of socio-economic systems~\citep{braun,mon78}. For 
example, our use of the logistic equation and the 
power-law correlation function in the phase plot was equally
effective in modelling the growth of companies~\citep{akr10}. 
This analogy between national economies and companies is of 
interest because studies point to universal mechanisms that 
underlie the economic dynamics of countries and
companies~\citep{sabhlmss96,lacms98}.
This commonality can help in understanding the dynamics of 
large companies, whose stock values can grow 
to the scale of national economies. On this point, we note that 
major stock indices of the six countries in our study show as 
much regularity~\citep{kvr20} as the GDP and trade growth of 
the same countries. 

\begin{acknowledgments}
We thank A. R. Dhakulkar and N. Sarkar. Comments from
J. Mulherkar, A. Parikh and M. Tiwari are appreciated.  
\end{acknowledgments}

\bibliography{kr0822arX}
\end{document}